\documentclass[prc,floatfix, twocolumn,showpacs, showkeys, preprintnumbers, nofootinbib, superscriptaddress,10pt]{revtex4-1}
\usepackage[utf8]{inputenc}
\usepackage[sort&compress]{natbib}
\usepackage{ulem}
\usepackage{bm,bbm}
\usepackage{times}
\usepackage{amssymb,amsbsy,amsmath,amsfonts}
\usepackage{graphicx}
\usepackage{float}
\usepackage[dvipsnames,svgnames,x11names]{xcolor}
\usepackage{morefloats}
\usepackage{srcltx}
\usepackage{slashed}
\usepackage{subfig}
\usepackage{multirow}
\usepackage{verbatim}
\usepackage{hyperref}
\usepackage{tabularx}
\usepackage{braket}
\usepackage{lipsum}
\usepackage{overpic}
\usepackage{caption}

\usepackage[commandnameprefix=ifneeded]{changes}
\definechangesauthor[name={XIAO Yang},color=BrickRed]{XY}
\usepackage{todonotes}
\setcommentmarkup{\todo[color={authorcolor!20},size=\scriptsize]{#3: #1}}

\makeatletter
\DeclareRobustCommand{\cev}[1]{%
  {\mathpalette\do@cev{#1}}%
}
\newcommand{\do@cev}[2]{%
  \vbox{\offinterlineskip
    \sbox\z@{$\m@th#1 x$}%
    \ialign{##\cr
      \hidewidth\reflectbox{$\m@th#1\vec{}\mkern4mu$}\hidewidth\cr
      \noalign{\kern-\ht\z@}
      $\m@th#1#2$\cr
    }%
  }%
}
\makeatother

\DeclareUnicodeCharacter{3000}{HEREHEREHERE}

\begin{document}

\title{ Effect of a repulsive three-body  interaction on the $DD^{(*)}K$ molecule }

\author{Ya-Wen Pan}
 \affiliation{School of Physics, Beihang University, Beijing 102206, China}
 \affiliation{Research Center for Nuclear Physics (RCNP), Ibaraki, Osaka 567-0047, Japan}

 \author{Jun-Xu Lu}
 \affiliation{School of Physics, Beihang University, Beijing 102206, China}

\author{Emiko Hiyama}
\affiliation{Department of Physics, Tohoku University, Sendai 980-8578, Japan} 
\affiliation{RIKEN Nishina Center for Accelerator-Based Science, 2-1, Hirosawa, Wako, Saitama 351-0198, Japan}

 \author{Li-Sheng Geng}
 \email{lisheng.geng@buaa.edu.cn}
 \affiliation{School of Physics,  Beihang University, Beijing 102206, China}
 \affiliation{Sino-French Carbon Neutrality Research Center, \'Ecole Centrale de P\'ekin/School of General Engineering, Beihang University, Beijing 100191, China}
 \affiliation{Peng Huanwu Collaborative Center for Research and Education, Beihang University, Beijing 100191, China}
 \affiliation{Beijing Key Laboratory of Advanced Nuclear Materials and Physics, Beihang University, Beijing 102206, China }
 \affiliation{Southern Center for Nuclear-Science Theory (SCNT), Institute of Modern Physics, Chinese Academy of Sciences, Huizhou 516000, China}

 \author{Atsushi Hosaka}\email{hosaka@rcnp.osaka-u.ac.jp}
 \affiliation{Research Center for Nuclear Physics (RCNP), Ibaraki, Osaka 567-0047, Japan}
 \affiliation{Advanced Science Research Center, Japan Atomic Energy Agency, Tokai, Ibaraki 319-1195, Japan}

\begin{abstract}
The hadronic molecular picture of the observed exotic states has inspired numerous investigations into few-body systems. Recently, the lattice effective field theory studied the effect of a three-body interaction on the binding energy of the $DD^{*}K$ system, revealing an intriguing phenomenon in the binding energy. This work uses the Gaussian expansion method to explore the underlying physics. Our results show that as the repulsive three-body interaction strengthens, the spatial size of the $DD^{(*)}K$ bound state gradually increases. Further enhancement of the three-body interaction causes the $DD^{(*)}K$ three-body bound state to break into a $D^{(*)}K$ two-body bound state, accompanied by a distant $D$ meson. 
The identical nature of the two $D$ mesons leads to the fact that the $DDK$ system consistently resembles an isosceles triangle-shaped spatial configuration.

\end{abstract}

\maketitle

\section{Introduction}\label{Intro}

The traditional quark model classifies hadrons into two categories: mesons as $q\Bar{q}$ bound states and baryons as $qqq$ bound states. 
However, Quantum Chromodynamics (QCD), the fundamental theory of the strong interaction, allows for more complex configurations, such as $qq\Bar{q}\Bar{q}$ mesons and $qqqq\Bar{q}$ baryons. These states, which the traditional quark model cannot explain, are often referred to as exotic states. 
In 2003, the Belle collaboration discovered a charmonium-like state $X(3872)$~\cite{Belle:2003nnu}, later confirmed by many other collaborations~\cite{BaBar:2004iez, CDF:2003cab, D0:2004zmu, CMS:2013fpt, LHCb:2011zzp, BESIII:2013fnz}. Its mass deviation from the quark model predictions~\cite{Godfrey:1985xj} and the observed isospin violation~\cite{Belle:2005lfc, BaBar:2010wfc, BESIII:2019qvy, LHCb:2022jez} suggest that it is unlikely a pure $c\Bar{c}$ charmonium state. Recent studies proposed that $X(3872)$ might be a mixture of two configurations, a $\Bar{D}^*D$ molecular configuration and a $\chi_{c1}(2P)$ charmonium configuration~\cite{Ortega:2009hj, Yamaguchi:2019vea, Duan:2021alw, Kinugawa:2023fbf, Deng:2023mza, Song:2023pdq, Takeuchi:2021cnp}, with the molecular component contributing more than 80\% to its wavefunction~\cite{BESIII:2023hml, Esposito:2021vhu, Baru:2021ldu, Li:2024pfg, Song:2023pdq}.  
$D_{s0}^*(2317)$, first discovered by the BaBar collaboration in 2003~\cite{BaBar:2003oey} and later confirmed by the CLEO~\cite{CLEO:2003ggt}, Belle~\cite{Belle:2003guh}, and BESIII~\cite{BESIII:2017vdm} collaborations, is also believed to have a significant $DK$ molecular component, contributing  more than 70\% to its wave function~\cite{MartinezTorres:2014kpc, Albaladejo:2018mhb, Yang:2021tvc, Guo:2023wkv, Gil-Dominguez:2023huq}.
In 2020, the LHCb collaboration discovered a narrow doubly charmed tetraquark state, $T_{cc}(3875)^*$, which lies about 350 keV below the $D^{*+}D^0$ threshold ~\cite{LHCb:2021vvq}. Recent compositeness analyses~\cite{Albaladejo:2021vln,Du:2021zzh,Kinugawa:2023fbf,Wang:2023ovj,Dai:2023kwv,Dai:2023cyo} suggested that $T_{cc}(3875)^*$ is predominantly a $DD^*$ molecular state.

Based on the molecular picture for $X(3872)$, $D_{s0}^*(2317)$, and $T_{cc}(3875)^+$, one can obtain the interactions between the respective hadron pairs, which has inspired numerous theoretical investigations into three-body systems composed of $D^{(*)}$ and $K^{(*)}$ mesons~\cite{Wu:2022ftm, Liu:2024uxn, MartinezTorres:2018zbl, Pang:2020pkl, Wu:2019vsy, Wu:2020rdg, Wu:2020job, Zhang:2024yfj, Debastiani:2017vhv, MartinezTorres:2012jr, Wei:2022jgc, Ma:2017ery, Tan:2024omp, Ren:2018pcd, Ikeno:2022jbb}. 
Ref.~\cite{Wu:2019vsy} predicted a $DDK$ bound state with a binding energy of about 70 MeV in the Gaussian expansion method (GEM), consistent with the 90 MeV from the chiral Faddeev equation~\cite{ MartinezTorres:2018zbl} and 70 MeV from finite volume~\cite{Pang:2020pkl}. The $D\Bar{D}K$ system was found to have a binding energy of about 45 MeV in the GEM~\cite{Wu:2020job} and fixed center approach~\cite{Wei:2022jgc}. The $DD^*K$ system has a binding energy of about 50 MeV~\cite{Ma:2017ery}. However, in Ref.~\cite{Tan:2024omp}, the authors found that the binding energy of the $DD^*K$ molecular state is very small, only 0.8 MeV, based on the quark level calculation. Additionally, bound states such as $D\Bar{D}^*K$~\cite{Wu:2020job, Ma:2017ery, Ren:2018pcd}, $D^* D^* \Bar{K}^*$~\cite{Ikeno:2022jbb}, $D^{(*)}D^{(*)}D^{(*)}$~\cite{Wu:2021kbu,  Luo:2021ggs, Bayar:2022bnc}, and $D^{(*)}D^{(*)}\Bar{D}^{(*)}$~\cite{Tan:2024omp, Valderrama:2018sap} have also been predicted.

Recently, the effect of a three-body interaction on the $DD^*K$ system was studied by the lattice effective field theory in Ref.~\cite{Zhang:2024yfj}. Without the three-body interaction, the $DD^*K$ system has a binding energy of about 79 MeV. The binding energy of the $DD^*K$ decreases as the strength of the three-body repulsive interaction increases and asymptotically approaches about 45 MeV when the repulsion is infinitely strong. The authors concluded that the three-body system remains bound even with an infinitely strong repulsive three-body interaction.

Interestingly, this binding energy of 45 MeV matches the binding energy of the $D_{s0}^*(2317)$ and $D_{s1}(2460)$ states as $DK$ and $D^*K$ molecular states, respectively. This raises an intriguing question: is there an underlying physics connecting these two numbers?  In this work, we employ the Gaussian expansion method to study the effect of the three-body interaction on the $DD^{(*)}K$ systems. This approach enables precise calculation of binding energies and provides valuable insights into the spatial configurations of the systems studied.

This article is organized as follows. In Sec.~\ref{frame}, we briefly explain how the GEM is used to solve the three-body Schr\"{o}dinger equation and the two-body and three-body interactions we employ. The results are presented and discussed in Sec.~\ref{re-dis}, followed by a summary in Sec.~\ref{Con}.

\section{Theoretical Framework}
\label{frame}

\begin{figure}[!htbp]
  \centering
  \begin{overpic}[scale=0.63]{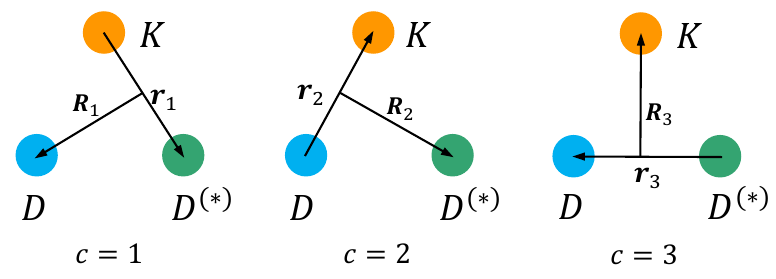}
  \end{overpic}
   \captionsetup{justification=raggedright, singlelinecheck=false}
  \caption{ Three Jacobi channels of the $DD^{(*)}K$ system. Channels $c=1,2$ should be symmetrized for the $DDK$ system.}
  \label{jaco}
\end{figure}

The three-body Schr\"{o}dinger equation is solved by the  Gaussian Expansion Method (GEM), which has been widely applied to study few-body systems~\cite{Hiyama:2003cu,Pan:2024ple,  Pan:2023zkl, Richard:2016eis, Yoshida:2015tia, Pan:2022xxz}:
\begin{equation}
    [T+\sum_{1=i<j}^{3}V(r_{ij}) + V_{123} -E]\Psi_{JM}=0\,,
\end{equation}
where $T$ is the kinetic energy, $V(r_{ij})$ are the two-body potentials between particles $i$ and $j$, $V_{123}$ is the three-body potential, and $\Psi_{JM}$ is the total wave function of the three-body system with $J$ the total angular momentum, which can be expressed as the sum of the wave functions of the three Jacobi channels in Fig.~\ref{jaco}:
\begin{equation}
    \Psi_{JM}=\sum_{c,\alpha} \mathcal{S}_{D} \Phi_{JM,\alpha}^{c}(\pmb{r}_{c},\pmb{R}_{c})\quad (c=1-3)\,,
\end{equation}
where the lower index $c$ denotes the three Jacobi channels. $\mathcal{S}_{D}$ is the symmetrization operator for the two channels ($c=1,2$) in the study of the $DDK$ system, required by the Bose-Einstein statistics for the two identical particles $D$.  
The wave functions of each Jacobi channel, denoted by $\Phi_{JM,\alpha}^{c}(\pmb{r}_{c},\pmb{R}_{c})$, are given by
\begin{equation}
    \Phi_{JM,\alpha}^{c}(\pmb{r}_{c},\pmb{R}_{c}) = C_{c,\alpha} \left\{\left[\phi^{G}_{n_{c}l_{c}}(\pmb{r}_{c})\psi^{G}_{N_{c}L_{c}}(\pmb{R}_{c})\right]_{\Lambda}\otimes \Omega_{S}\right\}_{JM}\,,
\end{equation}
where $C_{c,\alpha}$ are the expansion coefficients and $\alpha$ represents a set of labels $\{n,N,l, L, \Lambda, S\}$. In this notation, $l$ and $L$ are the orbital angular momenta associated with each channel's coordinates $r$ and $R$, and $\Lambda$ is the total orbital angular momentum. $\Omega_{S}$ is the spin wave function with $S$ the total spin angular momentum, which can be neglected in this work.
The functions $\phi^{G}_{n_{c}l_{c}}$ and $\psi^{G}_{N_{c}L_{c}}$ are the spatial wave functions that can be expanded in terms of the Gaussian functions of the Jacobi coordinates $\pmb{r}$, $\pmb{R}$:
\begin{equation}
    \begin{aligned}
    &\phi^{G}_{nlm}(\pmb{r})=\phi^{G}_{nl}(r)Y_{lm}(\hat{\pmb{r}})\,, \phi^{G}_{nl}(r)=N_{nl} r^{l} e^{-\nu_{n}r^{2}}\,, \\
    &\psi^{G}_{NLM'}(\pmb{R})=\psi^{G}_{NL}(r)Y_{LM'}(\hat{\pmb{R}})\,, \psi^{G}_{NL}(R)=N_{NL} R^{L} e^{-\lambda_{N}R^{2}}\,,
    \end{aligned}
\end{equation}
where $N_{nl}$ and $N_{NL}$ are the normalization constants.
The parameters $\nu_{n}$ and $\lambda_{N}$ are given by
\begin{equation}
   \begin{aligned}
    &\nu_{n}=1/r^{2}_{n},~~ r_{n}=r_{1}a^{n-1},~~~~~~~~~~(n=1, 2, ..., n_{max})\\
    &\lambda_{N}=1/R^{2}_{N},~~ R_{N}=R_{1}A^{N-1},~~(N=1, 2, ..., N_{max})
   \end{aligned}
\end{equation}
where $\{n_{max}, r_{1}, a~ \mbox{or}~r_{max} \}$ and $\{N_{max}, R_{1},  A~ \mbox{or}~ R_{max} \}$ are Gaussian basis parameters. In this work, we set $n_{max}=40$, $N_{max}=30$, $r_{min}=R_{min}=0.01$ fm, and $r_{max}=R_{max}=50$ fm.
Further optimization of the Gaussian parameters does not change the results. 
For a detailed calculation of the $DD^* K$ system by GEM, refer to Refs.~\cite{Wu:2019vsy,Wu:2020rdg,Wu:2020job}.

Since this work focuses on the effect of the three-body interaction, we employ the same two-body potentials as those in Refs.~\cite{Wu:2019vsy,Wu:2020rdg,Wu:2020job}. The $DD^{(*)}$ potential is described by the one-boson exchange model with a monopole form factor $\frac{\Lambda^2 - m^2}{\Lambda^2 - q^2}$. The cutoff $\Lambda$ in the form factor was determined by reproducing the mass of $X(3872)$ as a $D\Bar{D}^*$ bound state with a binding energy of 4 MeV, resulting in  $\Lambda = 1.01$ GeV.

The leading order (LO) $D^{(*)}K$ potential is described by the Weinberg-Tomozawa (WT) term. After the Fourier transformation with a Gaussian regulator, the potential in coordinate space reads 
\begin{equation}
    V_{D^{(*)}K}(r) = C_L e^{-(r/b)^2}\,,
\end{equation}
where $b$ is the effective range, and $C_L$ is a free parameter determined by reproducing the mass of $D_{s0}^*(2317)$ as a $DK$ molecule. In this work, we set $b=1$ fm and $C_L=-320.1$ MeV~\cite{Wu:2019vsy}.

For the three-body potential, we follow Ref.~\cite{Zhang:2024yfj}, where the LO three-body Lagrangian is given by
\begin{equation}
    \mathcal{L}_{DD^*K} = c_3\,\mathrm{Tr}\left[ H\mathcal{D}_{\mu}H^{\dagger}H\mathcal{D}^{\mu}H^{\dagger} \right ]\,,
\end{equation}
with $c_3$ a free parameter describing the strength of the three-body potential and $H$ the superfield of the charmed meson doublet. The covariant derivative is defined as 
\begin{equation}
    \mathcal{D}_{\mu} = \vec{\partial}_{\mu} - \cev{\partial}_{\mu} + (u^{\dagger}\partial_{\mu}u + u\partial_{\mu}u^{\dagger})/2\,,
\end{equation}
where $ u = \mathrm{exp}(\frac{i\phi}{\sqrt{2}f_{\pi}}) $ with $\phi$ representing the pseudoscalar meson octet,
\begin{equation}
    \phi = \begin{pmatrix}
 \frac{\pi^0}{\sqrt{2}}+\frac{\eta}{\sqrt{6}} & \pi^+ & K^+\\
 \pi^- & -\frac{\pi^0}{\sqrt{2}}+\frac{\eta}{\sqrt{6}} & K^0\\
  K^-& K^0 &-\frac{2\eta}{\sqrt{6}}
\end{pmatrix}\,.
\end{equation}
Based on the Lagrangian described above, the LO $S$-wave  three-body potential is
\begin{equation}
    V_{DD^*K} = \frac{c_3}{4f_{\pi}^2} (p_1+p'_1 +p_2+p'_2)(p_3+p'_3)\,.
\end{equation}
where $p_{1,2,3}$ and $p'_{1,2,3}$ are the four-momentum of the incoming and outgoing particles. With non-relativistic approximations, the three-body potential simplifies to a contact potential. By introducing the same regulator as employed in the $D^{(*)}K$ potential, the potential in coordinate space is given by
\begin{equation}
    V_{DD^*K}(r,R) = C \frac{e^{-r^2/a_1}}{\pi^{3/2}a_1^3}\frac{e^{-R^2/a_2}}{\pi^{3/2}a_2^3}\,,
\end{equation}
where $r$ and $R$ are the Jacobi coordinates illustrated in Fig.~\ref{jaco}. $a_1$ and $a_2$ are the effective ranges, and $C$ is a free parameter encapsulating all constants to describe the strength of the three-body potential. 
Due to the lack of sufficient information on three-hadron interactions, we consider the strength of the three-body potential to vary within the range $(0,+\infty)$ MeV.
This study focuses exclusively on the repulsive three-body potential, as an attractive one would trivially lead to a more bound $DD^*K$ system. 
Given the exploratory nature of this work, the effective ranges involved in the three-body potential are set to $a_1 = 0.5$ fm and $a_2 = 0.4$ fm. Notably, we have verified varying the effective range does not alter the qualitative conclusions of our study.

\section{Results and Discussions}\label{re-dis}

\begin{figure}[!htbp]
  \centering
  \begin{overpic}[scale=0.4]{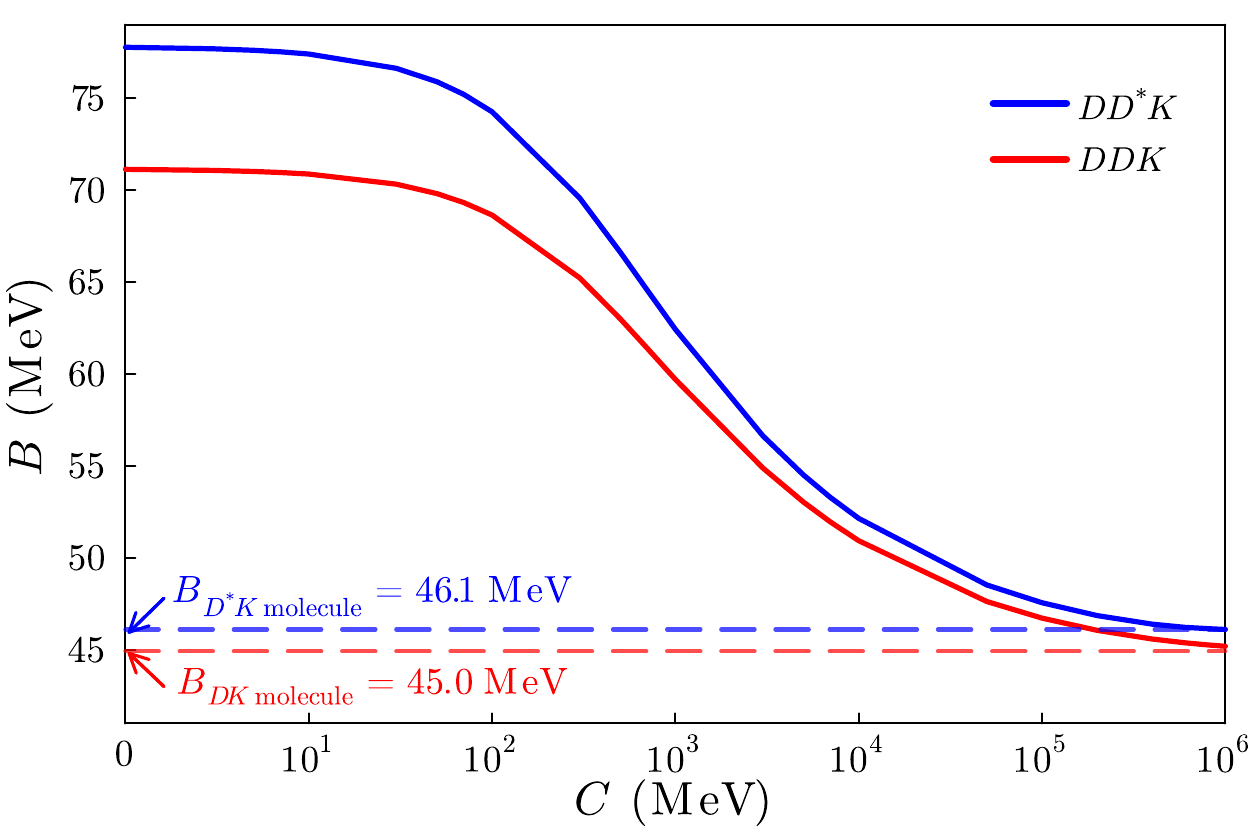}
  \end{overpic}
   \captionsetup{justification=raggedright, singlelinecheck=false}
  \caption{Binding energies of the $DD^{(*)}K$ systems as a function of the strength $C$ of the three-body potential.  \label{Bt}}
\end{figure}

Without the three-body potential, the $DD^*K$ system has a binding energy of 77.8 MeV, consistent with the 79.1 MeV in Ref.~\cite{Zhang:2024yfj}. 
Similarly, the $DDK$ system has a binding energy of 71.2 MeV, in agreement with the results of Ref.~\cite{Wu:2019vsy}.
As the strength of the three-body potential increases, the binding energies decrease, as shown in Fig.~\ref{Bt}, consistent with the finding of Ref.~\cite{Zhang:2024yfj}. These trends in the binding energies of the $DD^{(*)}K$ systems can be readily explained by the repulsive nature of the three-body potential.

An interesting phenomenon shown in Fig.~\ref{Bt} is that as the strength of the three-body potential approaches infinity, the binding energies of the $DD^{(*)}K$ three-body systems asymptotically approach those of the $D^{(*)}K$ two-body systems. Such a phenomenon also was found in Ref.~\cite{Zhang:2024yfj}. However, the authors didn't discuss it further. To understand the underlying physics, we investigate the spatial configuration of the $DD^{(*)}K$ three-body systems by calculating the distance between each hadron pair. This distance is related to the root-mean-square (rms) radii of Jacobi coordinate $r$, specifically, $\sqrt{\langle \Psi_{J}^{Total} |r^2_i| \Psi_{J}^{Total} \rangle}$, which we  abbreviate as $\langle r \rangle$ hereinafter.

\begin{figure*}[!htbp]
  \centering
  \begin{overpic}[scale=0.6]{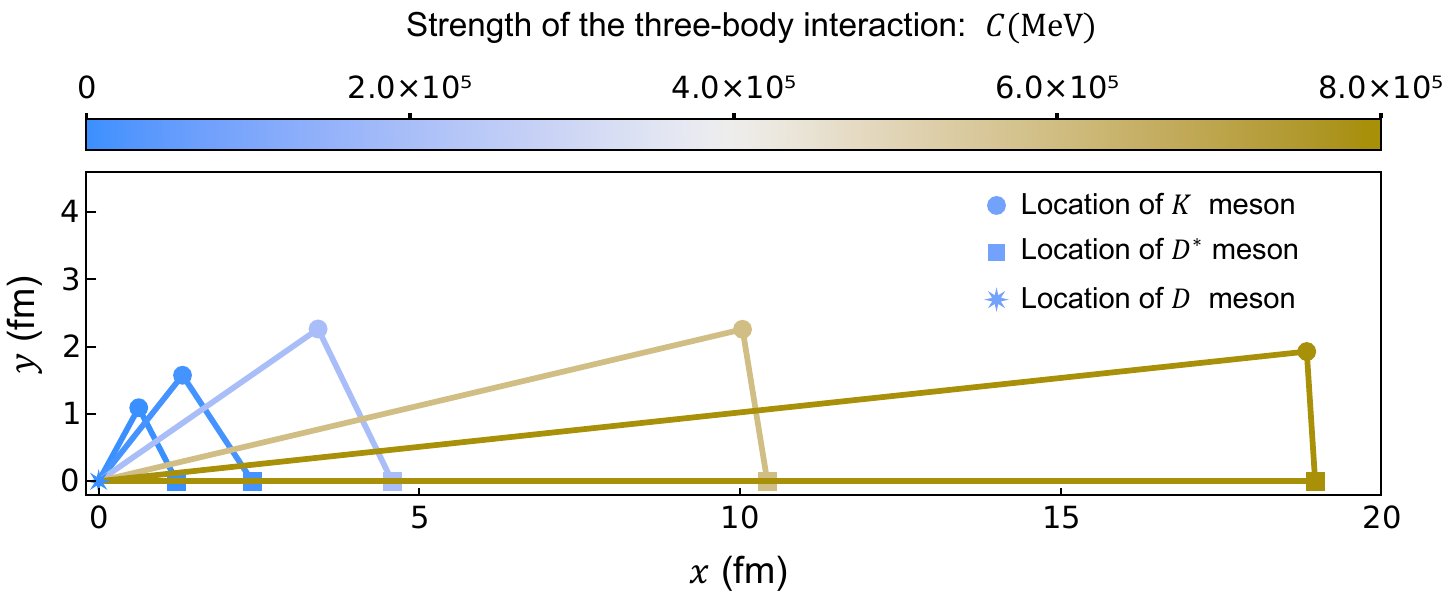}
  \end{overpic}
   \captionsetup{justification=raggedright, singlelinecheck=false}
  \caption{Spatial configuration of the $DD^*K$ system as a function of the strength of the three-body potential, depicted in a two-dimensional $x$-$y$ plane. The $D$ meson is fixed at the origin $(0,0)$. The length of the line segment connecting any two particles represents the corresponding rms radius $\langle  r \rangle$, with the color indicating the strength of the three-body potential. \label{shape-DDsK}}
\end{figure*}

Next, we analyze each case individually. First, for the $DD^{*}K$ system, we plot its spatial configuration as a function of the strength of the three-body potential in the two-dimensional $x$-$y$  plane, with the scale in fm, as shown in Fig.~\ref{shape-DDsK}. 
There are two stages in the evolution. In the first stage, where $C \lesssim 2\times 10^{5}$ MeV, the distances between each hadron pair increase as the strength of the three-body potential increases, which is in accordance with intuition. 
In the second stage, where $C \gtrsim 2\times 10^{5}$ MeV, as the strength of the three-body potential continues to increase, the distances between $DD^*$ and between $DK$ keep increasing, while the distance between $D^* K$  begins to decrease. When the strength of the three-body potential becomes infinite, the distance between $D^* K$ in the $DD^{*}K$ three-body system is 1.27 fm, corresponding to the characteristic size of a $D^* K$ two-body bound state. These results suggest that as the three-body potential strengthens, the size of the $DD^{*}K$ bound state gradually increases. The continued strengthening of the potential eventually breaks apart the $DD^{*}K$ three-body bound state, resulting in a $D^{*}K$ two-body bound state with a distant $D$ meson.

The $DD^{*}K$ bound state can be viewed as a superposition of three configurations: $[D^*K]D$, $[DK]D^*$, and $[DD^*]K$, corresponding to the three Jacobi channels shown in Fig.~\ref{jaco}. The notation $[D^*K]D$ refers to a $D^*K$ cluster with a $D$ meson. We then study the probabilities of each configuration, which are reflected in the probabilities of the Jacobi channels, denoted as $\mathcal{P}_c$.
Sine the attraction of the $D^{(*)}K$ potential is much stronger than that of the $D^{*}D$ potential, the contribution from channel $c=3$ (the $[DD^*]K$ configuration) can be neglected. Thus, in the following discussion, we focus only on channels $c=1$ and $c=2$. 
It is important to note that the basis functions are not orthogonal to each other, so we cannot directly obtain the exact probability of each channel, which is calculated by $\langle\Psi^{c}_{J}|\Psi^{c}_{J}\rangle$. However, the expectation values of the two-body potentials can provide an estimate of the probability to some extent, i.e., $\mathcal{P}_1 \sim  \langle \Psi_{J}^{Total} | V_{D^* K} | \Psi_{J}^{Total} \rangle$, and $\mathcal{P}_2 \sim \langle \Psi_{J}^{Total} | V_{D K} | \Psi_{J}^{Total} \rangle$.

Table~\ref{re-DDsK} collects the binding energies, expectation values of the $D^{(*)}K$ potentials, and rms radii of the $DD^*K$ three-body system. As the strength of the three-body potential increases, one can observe several trends.
In the first stage ($C \lesssim 2\times 10^{5}$ MeV), the distance between the $D^*$ and $K$ mesons increases, whereas in the second stage ($C \gtrsim 2\times 10^{5}$ MeV), this distance begins to decrease. Meanwhile, the distance between the $D$ and $K$ mesons and between the $D$ and $D^*$ mesons continues to increase. 
Additionally, the probability of channel $c=1$ consistently grows, indicating that the $[D^*K]K$ configuration plays an increasingly dominant role. When the strength of the three-body potential becomes infinite, we find that $\mathcal{P}_1 = 100\%$, $\langle  r_{D^* K}  \rangle = 1.27$ fm, and $\langle  r_{D K}  \rangle \to \infty$. These results indicate that the $DD^*K$ three-body bound state has effectively transformed into a $D^*K$ two-body bound state, with the $D$ meson separated far away.

\begin{table}[!htbp]
    \centering
     \captionsetup{justification=raggedright, singlelinecheck=false}
    \caption{ Binding energies, expectation values, and rms radii of the $DD^*K$ three-body system.  Energies are in units of MeV, and radii are in units of fm. The values inside the brackets are calculated as $\langle  V_{D^{(*)} K}  \rangle /\langle V_{D^* K} + V_{D K} \rangle $. \label{re-DDsK}}
     \scalebox{0.94}{
    \begin{tabular}{ccccccccccccc}
    \hline\hline
     $C$ & $B_{DD^*K}$ & $\langle  V_{D^* K}  \rangle$~($\mathcal{P}_1$) & $\langle  V_{D K}  \rangle$~($\mathcal{P}_2$) & $\langle  r_{D^* K}  \rangle$ & $\langle  r_{D K}  \rangle$ \\\hline
    $0$ & $77.8$ &  $-100.5~(50.6\%)$ & $-98.1~(49.4\%)$ & $1.24$ & $1.26$\\
    $1\times 10^4$ & $52.1$ & $-72.5~(52.6\%)$ & $-65.4~(47.4\%)$ & $1.92$  & $2.05$\\
    $2\times 10^5$ & $46.9$ & $-79.4~(62.8\%)$ & $-47.1~(38.2\%)$ & $2.54$ & $4.01$ \\
    $6\times 10^5$ & $46.2$ & $-100.4~(80.0\%)$ & $-25.1~(20.0\%)$ & $2.29$ & $10.29$ \\
    $8\times 10^5$ & $46.1$ & $-111.8~(88.9\%)$ & $-13.9~(11.1\%)$ & $1.93$ & $18.93$ \\
    $\infty$ & $46.1$ & $-125.9~(100\%)$ & $ 0~(0\%)$ & $1.27$ & $\infty$ \\
    \hline\hline
    \end{tabular}}
\end{table}

In addition, we can qualitatively understand this phenomenon from the potential between the $K$ meson and the $DD^*$ cluster, i.e., $V_{K[DD^*]} = V_{D^* K} + V_{DK} + V_{123}$. 
Due to the many independent variables involved, plotting this potential intuitively and accurately is difficult, so we make two approximations.
First, the distance between the $D$ and $D^*$ mesons is chosen to be $\langle  r_{D D^*}  \rangle$, as obtained from the GEM results. We then set the center of mass of the $DD^*$ cluster at the origin in the $(x,y)$ plane. The positions of the $D$ and $D^*$ mesons are given by $(-\frac{m_{D^*}}{m_{D} + m_{D^*}}\langle r_{DD^*} \rangle ,0)$ and $(\frac{m_{D}}{m_{D} + m_{D^*}}\langle r_{DD^*} \rangle ,0)$, respectively. 
Second, for the same two-body potential, the binding energy of the $D^*K$ system is larger than that of the $DK$ system due to the larger reduced mass (or smaller kinetic energy) of the $D^*K$ system. Since the $D$ and $D^*$ mesons are fixed, we assume that the $D^*K$ potential is more attractive to account for the kinetic energy effect. This is reflected by  $V_{D^* K} =k V_{D K}$. We choose $k=1.3$ for illustrative purposes to make the effect more visible.


\begin{figure*}[!htbp]
  \centering
  \begin{overpic}[scale=0.75]{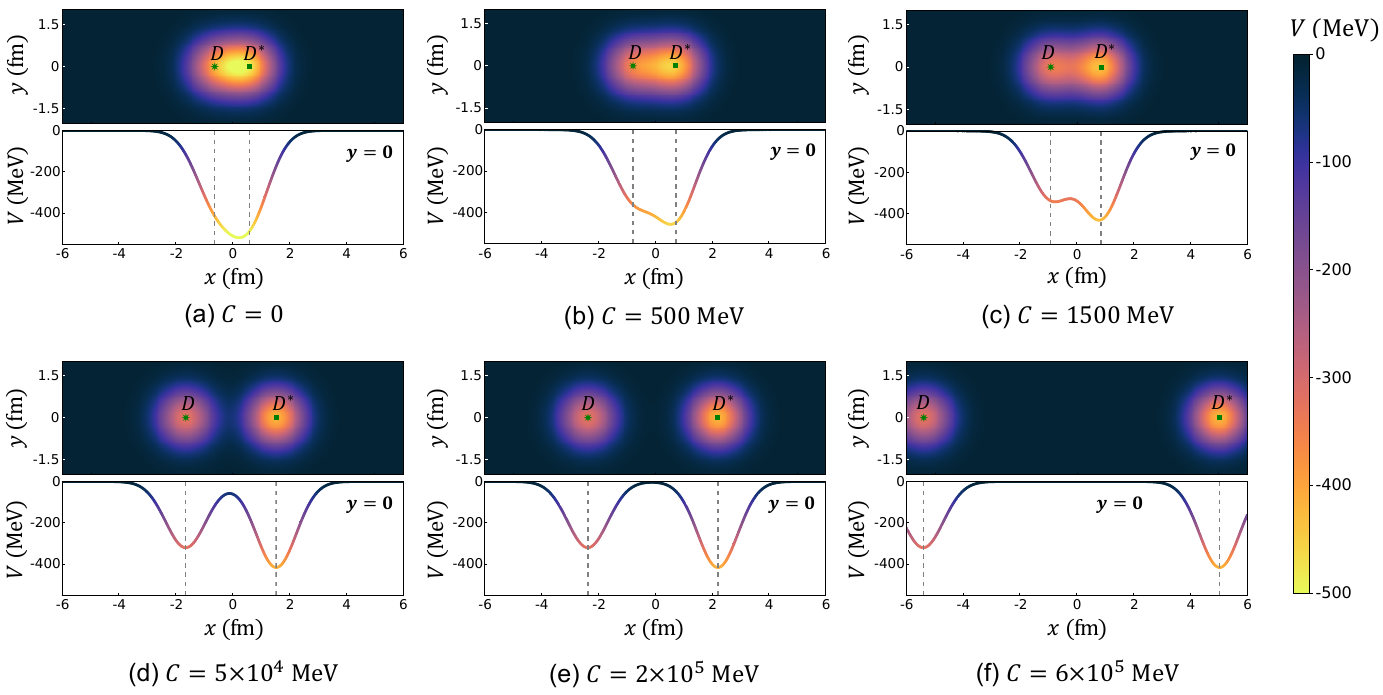}
  \end{overpic}
  \caption{The effective potential between the $K$ meson and the $DD^*$ cluster. The lower panels in each sub-figure are for $y=0$. \label{V-DDsK-heat}}
\end{figure*}

With the two approximations mentioned above, the effective potentials~\footnote{Since the potential is not exact due to the approximations, we refer to it as the ``effective potential''} between the $K$ meson and the $DD^*$ cluster are shown in Fig.~\ref{V-DDsK-heat}. Without the three-body potential, $V_{K[DD^*]}$ resembles a cone-shaped potential well, which is nearly symmetric between the $D$ and $D^*$ mesons, with the lowest point closer to the $D^*$ meson. 
This potential attracts the $K$ meson closer to the $D^*$ meson, i.e., the distance between $D^*K$ being slightly smaller than that of $DK$, which is consistent with the results presented in Table~\ref{re-DDsK}. 
As the strength of the three-body potential increases, a saddle point gradually appears in the effective potential. The minimum of the potential is closer to the $D^*$ meson, indicating that the $[D^*K]D$ is the most stable configuration. When $C \lesssim 2\times 10^{5}$ MeV, the saddle point is negative, allowing for a high transition probability from the $[D^*K]D$ configuration to the $[DK]D^*$ configuration. This means both configurations contribute to the three-body bound state. Consequently, the superposition of these two configurations increases the size of the three-body bound state, as shown in Fig.~\ref{shape-DDsK}. 
When $C \gtrsim 2\times 10^{5}$ MeV, the potential splits into two parts, and the distance between the $D$ and $D^*$ mesons continues to grow as the potential strength increases, which makes the transition from the $[D^*K]D$ configuration to the $[DK]D^*$ configuration increasingly difficult. As a result, when the strength becomes infinite, the $DD^*K$ three-body system with a binding energy of 46.1 MeV effectively becomes a $D^*K$ two-body bound state with a distant $D$ meson.

Since the difference in the binding energies between the $D^*K$ and $DK$ two-body bound states is only 1 MeV, a configuration in which a $DK$ two-body bound state with a distant $D^*$ meson should also exist when the strength of the three-body potential becomes infinite. However, the Gaussian expansion method only identifies the configuration with the lowest energy, specifically, a $D^*K$ two-body bound state with a distant $D$ meson. Consequently,  this work does not explore the configuration involving a $DK$ two-body bound state with a distant $D^*$ meson.

\begin{figure}[!htbp]
  \centering
  \begin{overpic}[scale=0.35]{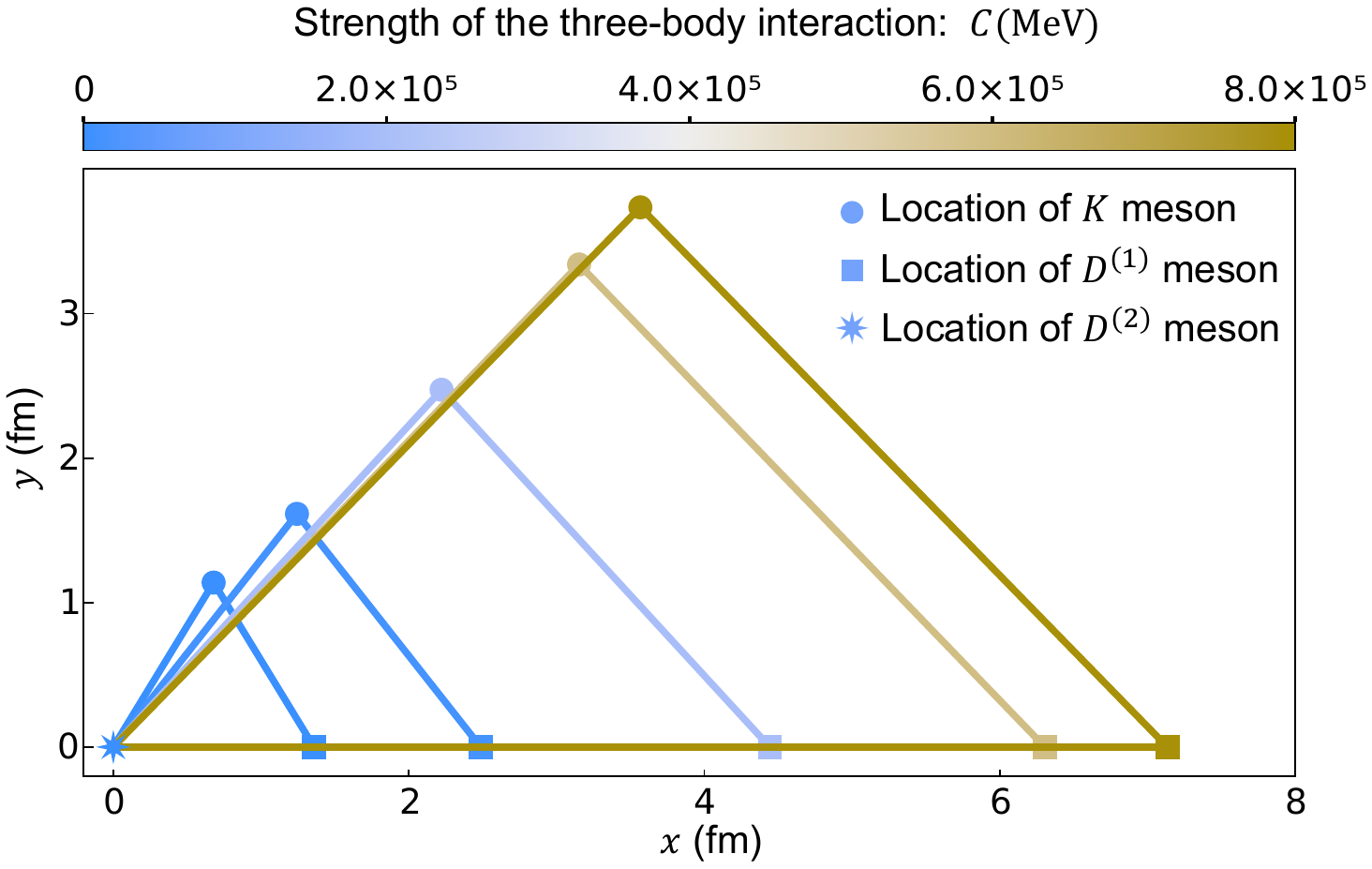}
  \end{overpic}
   \captionsetup{justification=raggedright, singlelinecheck=false}
  \caption{Spatial configuration of the $D^{(1)} D^{(2)} K$ system as a function of the strength of the three-body potential, depicted in a two-dimensional $x$-$y$ plane. The $D^{(2)}$ meson is fixed at the origin $(0,0)$. The length of the line segment connecting any two particles represents the corresponding rms radius  $\langle  r \rangle$, with the color indicating the strength of the three-body potential. \label{shape-DDK}}
\end{figure}

The results differ for the $DDK$ system, yet the underlying physics remains the same. We label the two identical $D$ mesons as $D^{(1)}$ and $D^{(2)}$ to clarify the discussion.  Fig.~\ref{shape-DDK} shows its spatial configuration in the same manner as the $DD^*K$ system. 
One can find that as the strength of the three-body potential increases, the size of the $D^{(1)}D^{(2)}K$ system continuously increases. This behavior is similar to the first stage of the $DD^*K$ system but differs from its second stage. 
The system's isosceles triangle shape is straightforward to understand due to the two identical $D$ mesons. Remarkably, when the strength of the three-body potential becomes infinite, the $DDK$ system with a binding energy of 45 MeV becomes very large.

Similarly, the $D^{(1)}D^{(2)}K$ bound state can be viewed as a superposition of the $[D^{(1)}K]D^{(2)}$ and $[D^{(2)}K]D^{(1)}$ configurations. Table~\ref{re-DDK} lists the binding energies, expectation values, and rms radii of the $D^{(1)} D^{(2)} K$ three-body system. It is important to note that the rms radii in Table~\ref{re-DDK} are calculated by the single Jacobi channel wave function, i.e., $\langle  r_{c}  \rangle_{c} = \sqrt{\langle \Psi_{J}^{c} |r^2_c| \Psi_{J}^{c} \rangle}$, and $\langle  R_{c}  \rangle_{c} = \sqrt{\langle \Psi_{J}^{c} |R^2_c| \Psi_{J}^{c} \rangle}$ where $R_c$ is the Jacobi coordinate depicted in Fig.~\ref{jaco}. 
Since the single Jacobi channel wave function is extracted from the total wave function, $\langle  r_{c}  \rangle_{c} $ does not represent the actual distance between the $D^{(1/2)}$ and $K$ mesons in the $D^{(1/2)}K$ cluster but serves as an indicator of its trend. Additionally, $\langle  R_{c}  \rangle_{c}$ calculated by the single Jacobi channel wave function reflects the distance between the $D$ meson and the $DK$ cluster.

\begin{figure*}[!htbp]
  \centering
  \begin{overpic}[scale=0.75]{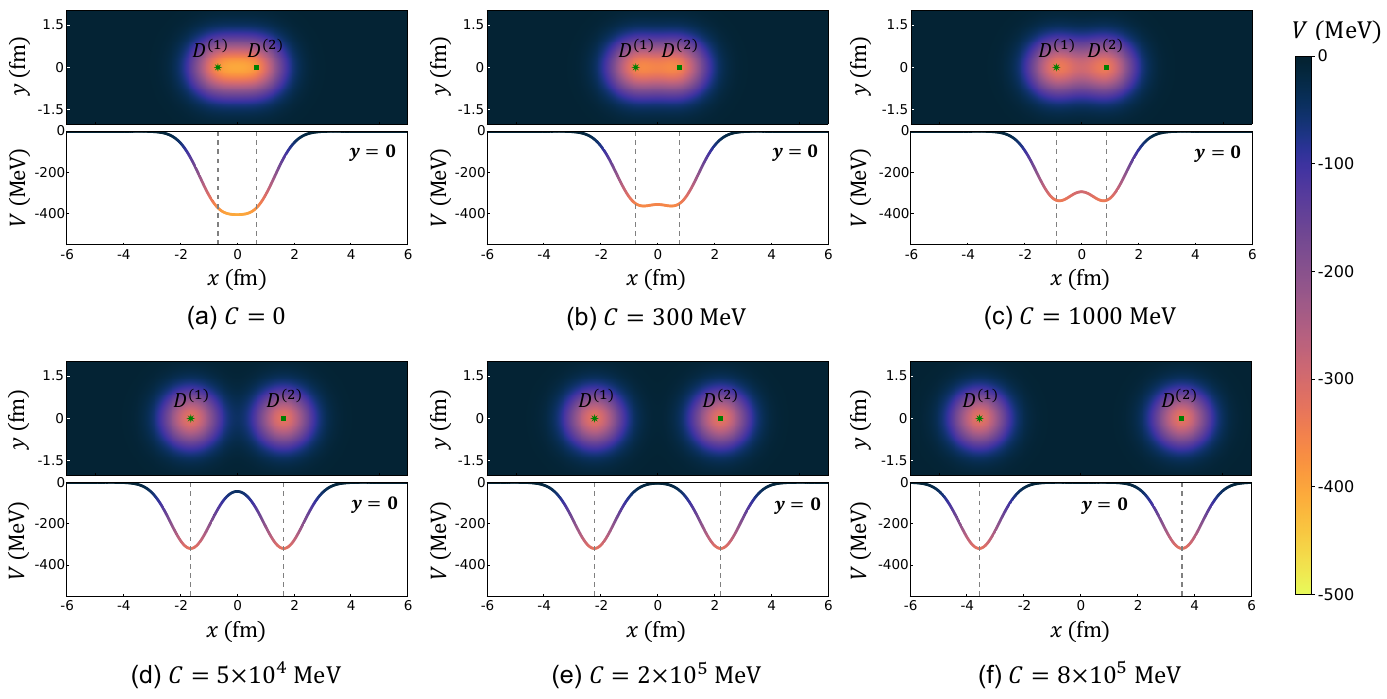}
  \end{overpic}
  \caption{The effective potential between the $K$ meson and the $DD$ cluster. The lower panels in each sub-figure are for $y=0$. \label{V-DDK-heat}}
\end{figure*}

\begin{table}[!htbp]
 \setlength{\tabcolsep}{7pt}
    \centering
     \captionsetup{justification=raggedright, singlelinecheck=false}
    \caption{Binding energies, expectation values, and rms radii of the $DDK$ three-body system. Since $\langle  V_{D^{(1)} K}  \rangle = $ $\langle  V_{D^{(2)} K}  \rangle$, $\langle  r_{D^{(1)} K}  \rangle_{1} = \langle  r_{D^{(2)} K}  \rangle_{2}$, and $\langle  R_{1}  \rangle_{1} = \langle  R_{2}  \rangle_{2}$, we provide only one value for each pair. Energies are in units of MeV, and radii are in units of fm. \label{re-DDK}}
     \scalebox{1.0}{
    \begin{tabular}{ccccccccccccc}
    \hline\hline
     $C$ & $B_{DDK}$ & $\langle  V_{D^{(1)} K}  \rangle $~($\mathcal{P}_1$) & $\langle  r_{D^{(1)} K}  \rangle_{1} $  & $\langle  R_{1}  \rangle_{1}$  \\\hline 
    $0$ & $71.2$ &  $-94.8$~($50\%$) & $1.02$  & $1.41$ \\
    $1\times 10^4$ & $50.9$ &  $-68.3$~($50\%$) & $1.45$  & $2.41$\\
    $2\times 10^5$ & $46.1$ &  $-62.4$~($50\%$) & $3.32$  & $4.01$\\
    $6\times 10^5$ & $45.4$ &  $-62.1$~($50\%$) & $2.36$  & $3.25$\\
    $8\times 10^5$ & $45.3$ &  $-62.0$~($50\%$) & $2.26$  & $3.37$\\
    $\infty$ & $45.0$ &  $-62.2$~($50\%$)  & $1.29$  & $\infty$ \\
    \hline\hline
    \end{tabular}}
\end{table}

om the behavior of $\langle  r_{D^{(1)} K}  \rangle_{1} $ in Table ~\ref{re-DDK}, one can find that as the strength of the three-body potential increases, the distance between the $D^{(1)}$ and $K$ mesons initially increases and then decreases. Meanwhile, the distance between the $D^{(2)}$ meson and the $D^{(1)}K$ cluster, as reflected by $\langle  R_{1}  \rangle_{1}$, shows a general increase~\footnote{ When $C \gtrsim 2\times 10^{5}$ MeV, $\langle  R_{1}  \rangle_{1}$ decreases slightly before increasing again. This is due to the influence of other channels on the single Jacobi channel wave function extracted from the total wave function.}. 
These trends indicate that as the strength of the three-body potential increases, the size of the $D^{(1)}D^{(2)}K$ bound state gradually increases. The continued increase eventually breaks apart the $D^{(1)}D^{(2)}K$ three-body bound state into a $D^{(1)}K$ two-body bound state with a distant $D^{(2)}$ meson, similar to what is observed in the $DD^*K$ system. 
From the perspective of $\langle  r_{D^{(2)} K}  \rangle_{2} $ and $\langle  R_{2}  \rangle_{2}$, the result corresponds to a $D^{(2)}K$ two-body bound state with a distant $D^{(1)}$ meson. 
Furthermore, the expectation values of the $D^{(1)}K$ and $D^{(2)}K$ two-body potentials indicate that,  regardless of the strength of the three-body potential, the probabilities of the $[D^{(1)}K]D^{(2)}$ and $[D^{(2)}K]D^{(1)}$ configurations remains equal. Consequently, the superposition of these two configurations causes the expression that the $D^{(1)}D^{(2)}K$ system continually expands, as shown in Fig.~\ref{shape-DDK}. This conclusion is further supported by the effective potential between the $K$ meson and $D^{(1)}D^{(2)}$ cluster, as shown in Fig.~\ref{V-DDK-heat}. 
With the three-body potential, there are two minima, indicating that both $[D^{(1)}K]D^{(2)}$ and $[D^{(2)}K]D^{(1)}$ are equally stable configurations.

We can conclude that the $DDK$ system exhibits two evolutionary stages as the strength of the three-body potential increases. In the first stage, the three-body system gradually expands, while in the second stage, it breaks into a $DK$ cluster with a distant $D$ meson.
The identical nature of the two $D$ mesons makes the $DDK$ system resemble an isosceles triangle.

\section{Summary}\label{Con}

In this work, we employed the Gaussian expansion method to investigate the effect of the three-body interactions on the $DD^{(*)}K$ bound states. The one boson exchange model provides the $DD^{(*)}$ potentials, while the $D^{(*)}K$ potentials are described by the Weinberg-Tomozawa term. The three-body interaction is formulated as a contact potential with a free parameter to describe its strength. Without the three-body potential, the $DD^{*}K$ and $DDK$ systems are bound with binding energies of 77.8 MeV and 71.2 MeV, respectively. As the strength of the three-body potential increases, the binding energies of the  $DD^{(*)}K$ bound states gradually decrease. In the limit where the strength becomes infinite, the three-body systems' binding energies approach those of the $D^{(*)}K$ two-body bound states.

The spatial configurations of the $DD^{(*)}K$ systems are characterized by the root mean square (rms) radii between each hadron pair in the three-body system. Without the three-body potential, the sizes of the $DD^{(*)}K$ bound states are 1 $\sim$ 2 fm, consistent with that of a typical molecular state. Introducing a three-body repulsive interaction produces distinct phenomena in the $DD^{*}K$ and $DDK$ systems. However, the same underlying physics governs both systems.

For the $DD^{*}K$ systems, as the strength of the three-body interaction increases, there are two stages in the evolution of its spatial configuration. In the first stage, the $DD^{*}K$ system gradually expands. In the second stage, the distance between the $D^*$ and $K$ meson begins to decrease, while the distance between the $D$ meson and the $D^*K$ cluster continues to increase.  When the strength of the three-body interaction becomes infinite, the distance between the $D^*$  and $K$ mesons reaches 1.27 fm, the same as the distance in the $D^*K$ two-body bound state. This indicates that, with an infinite three-body interaction strength, the $DD^{*}K$ bound state will break into a $D^*K$ two-body bound state with a distant $D$ meson. 
The reason is that the $DD^{*}K$ bound state can be viewed as a superposition of two configurations: a $D^{*}K$ cluster with a $D$ meson, denoted as $[D^{*}K]D$, and a $DK$ cluster with a  $D^{*}$ meson, denoted as $[DK]D^{*}$. Since the kinetic energy of the $D^{*}K$ cluster is smaller than that of the $DK$ cluster, the $[D^{*}K]D$ configurations are more stable. In the first stage, both configurations contribute significantly. However, in the second state, the transition from the $[D^{*}K]D$ to the $[DK]D^{*}$ configuration becomes increasingly difficult, with the $[D^{*}K]D$ configuration playing a more dominant role. When the strength of the three-body potential becomes infinite, the $[D^{*}K]D$ configuration accounts for 100\%.

The two stages for the $DDK$ system are similar to those in the $DD^{*}K$ system. When the three-body potential becomes infinitely strong, the $DDK$ bound state will break into a $DK$ two-body bound state with a distant $D$ meson. However, the identical nature of the two $D$ mesons results in a larger isosceles triangle-shaped spatial configuration.

The strength of the repulsive three-body interaction plays a crucial role in determining the binding energies and spatial configurations of three-body systems, making future experimental studies on three-hadron interactions essential for gaining deeper insights into their dynamics and properties.

\section{Acknowledgments}
L.S.G thanks Prof. Qian Wang for enlightening discussions. This work was partly supported by the National Key R\&D Program of China under Grant No.2023YFA1606703 and the National Natural Science Foundation of China under Grants No.12347113.
Y.W.P. acknowledges support from the China Scholarship Council scholarship and the Academic Excellence Foundation of BUAA for PhD Students.
A.H. is supported in part by the Grants-in-Aid for Scientific Research [Grant No. 21H04478(A), 24K07050(C)].
E.H. is supported in part by ERATO-JPMJER2304, Kakenhi-23K03378, and 20H00155.

\bibliography{Refs}

\end{document}